\documentclass[sigconf]{acmart}
\AtBeginDocument{%
  }

\setcopyright{cc}

\copyrightyear{2026}
\acmYear{2026}

\acmDOI{10.1145/3767308.3836019}

\acmConference[MM '26] {Proceedings of the 34th ACM International Conference on Multimedia}{November 10--14, 2026}{Rio de Janeiro, Brazil.}
\acmISBN{979-8-4007-2213-4/2026/11}

\setcctype{by}

\begin{document}

\title{Human-aware Design Generation: Adding 3D Humans into Graphic Designs}

\author{Zijin Hou}
\affiliation{%
  \institution{ShanghaiTech University}
  \city{Shanghai}
  \country{China}
}
\email{houzj2025@shanghaitech.edu.cn}
\orcid{0009-0001-3078-4964}

\author{Ying Cao}
\affiliation{%
  \institution{ShanghaiTech University}
  \city{Shanghai}
  \country{China}
}
\email{caoying59@gmail.com}
\orcid{0000-0002-9288-3167}
\authornote{Corresponding author.}

\begin{abstract}
Graphic designs nowadays predominantly contain human images, where people are posed and framed in a delicate way to guide viewers' visual exploration and evoke desired feelings. Despite recent progress in automatic graphic design generation, there has been no work on investigating the role of the human representations in design creation. In this paper, we seek to emphasize the importance of visual human representations in graphic design generation, by studying a novel task, dubbed as design-conditioned 3D human adding. To solve this task, we propose a model that adds 3D human representations into an existing partial design, by predicting their 3D poses, their 2D composition in the image containing them, as well as the spatial arrangement of the image in the input design, to generate a complete design.
Our experiments show that our model outperforms baselines by predicting the 3D human poses, 2D framing and human image layout that better harmonize with the other existing design elements, and generating higher-quality holistic designs. Our work is the first to explore connections between 3D human generation and 2D graphic design generation, which would inspire future research in leveraging 3D information to build more powerful generative models for graphic designs in the 2D domain.The code is available at
\href{https://github.com/GreyNails/Human-aware-Design-Generation-Adding-3D-Humans-into-Graphic-Designs}{GitHub}.
\end{abstract}



\begin{teaserfigure}
  \centering
  \includegraphics[width=\textwidth]{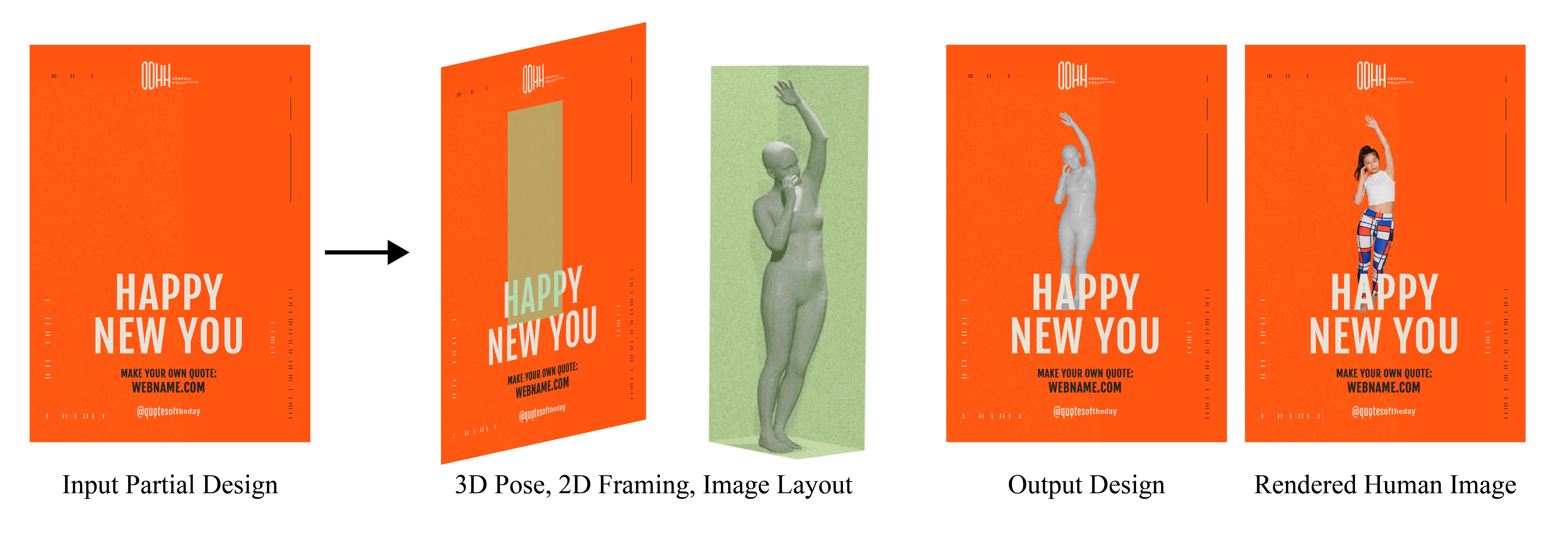}
  \vspace{-2.5em}
  \caption{
  Design-conditioned 3D human adding. Given an input design (1st column) with a set of elements, such as images, vector shapes and texts, our model adds a 3D human into the design (2nd column), by predicting its 3D pose and 2D framing, as well as the layout of the image containing the human in the input design (the 2D box on the design). The model results in a cohesive output design with the 3D human added (3th column). The 3D human representation can further be rendered into a realistic human image (4th column).
   }
  \label{fig:introduction}
\end{teaserfigure}

\maketitle

\section{Introduction}
\label{sec:intro}

\begin{figure}[t]
  \centering
  \includegraphics[width=0.9\linewidth]{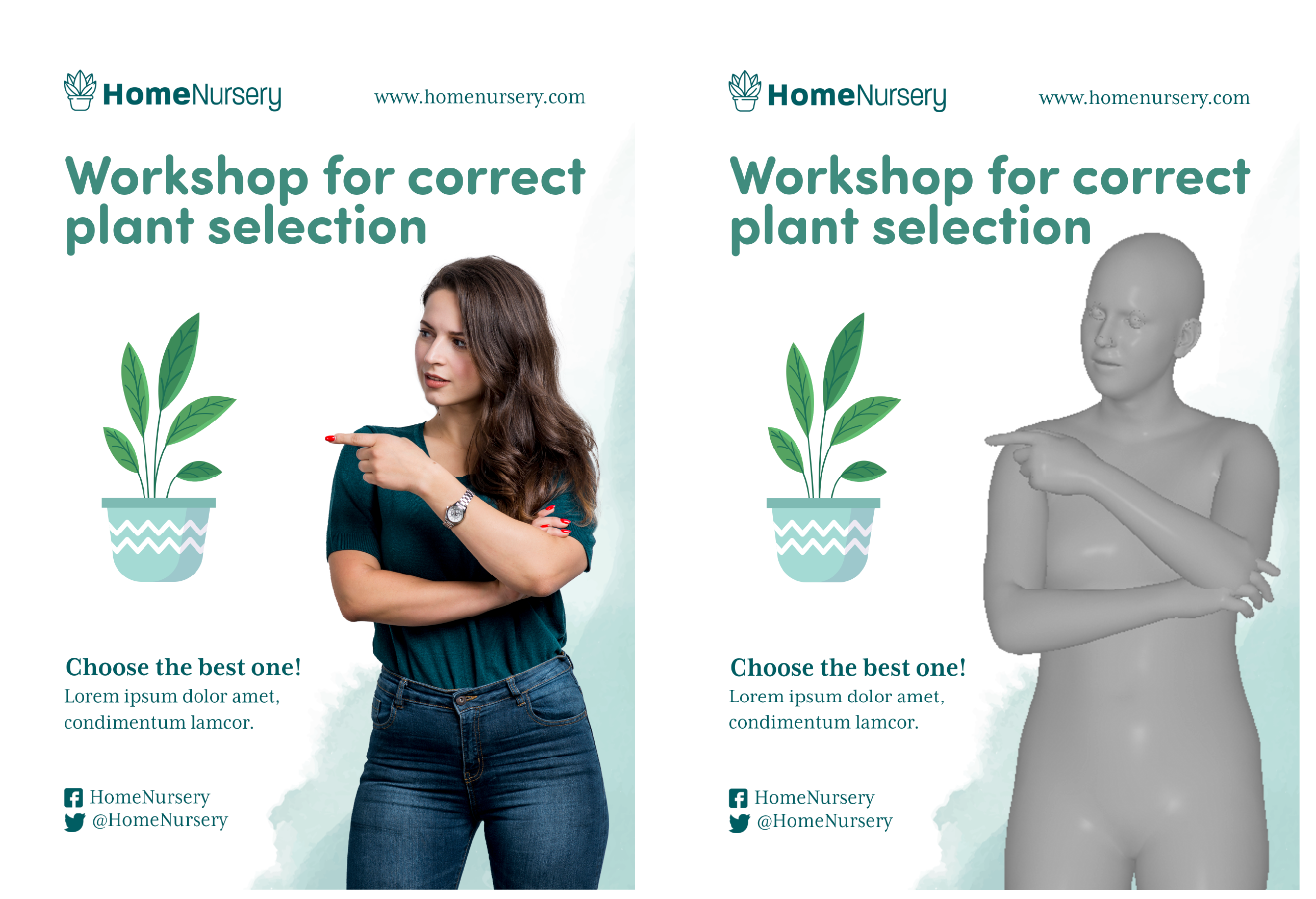}
  \caption{An example graphic design with a human image (left) and the design with the 3D representation of the human (right).}
  \label{fig:example}
\end{figure}

The use of human images is prevalent in graphic designs such as posters, magazines and advertisements. What poses people in an image take and how they are framed in the image play critical roles in affecting how viewers visually explore the entire design and what feelings are evoked. Consider the design in Figure~\ref{fig:example}
as an example. When looking at the design, one may first attend to the face of the woman, and then move the eyes to her hand near the face, Finally, the visual attention shifts to the left along the direction pointed by the finger, reaching the potted plant. This means that the pose of the woman can effectively direct one's visual flow throughout the design. Furthermore, the woman is framed in a medium full shot that starts from the top of the head and ends just below the waist, which conveys the feelings of ``strong'' and ``confident''. However, creating a human image with ideal posing and framing for a design is particularly challenging, requiring extensive design knowledge and experience, along with some creativity.

In recent years, there has been a growing interest in developing generative models for graphic design with some encouraging results.
However, the role of visual human representations has not yet been considered explicitly, which hampers the generation of designs where human images are employed effectively.    

In view of the issue above, in this paper, we take an initial step towards \textit{human-aware design generation}, by investigating a new task, \textit{design-conditioned 3D human adding}. As shown in Figure \ref{fig:introduction}, given an input partial design without a human image, the task aims to add 3D human representations into the input design to form a cohesive complete design, where the added human representations and the other elements are harmonious with each other. The generated 3D representations can further be rendered into a realistic human image. 
The purpose of generating 3D representations explicitly, instead of only 2D images, is to offer great controllability, which allows users to easily change the poses and composition of people in the rendered image through manipulating the generated 3D representations directly. Such controllability is imperative for the practical deployment of design generative models in the design workflow, since designers wish to have a higher degree of control over the generation process when working with the models.

We propose \textit{POFRAY}, \textbf{Po}se-\textbf{Fr}ame-L\textbf{ay}out, a model to solve this new task. POFRAY uses a partial layered design with a set of multimodal elements as a context, and add 3D humans into it by predicting what \textit{poses} they should take in 3D space, how they are \textit{framed} within the 2D image containing them, as well as how the image should be \textit{laid out} in the design. 
Specifically, POFRAY employs a multimodal encoder to process the input elements to extract design context features. The features are then used to condition two separate generators: a pose and framing generator, a layout generator. 
The pose and framing generator employs a causal Transformer to autoregressively generate a sequence of tokens that control the 3D poses of the added humans and their 2D framing in the image. The layout generator uses a bidirectional Transformer to generate the layout of the added human image in the input design. To train and evaluate our model, we construct \textit{3DHumanInDesign}, a benchmark of $\sim$10k layered graphic designs. For each design, we remove the human image
to form the input, and use the estimated human pose and framing parameters from the human image alongside the original layout of the the human image as the prediction targets.

Experimental results on our collected benchmark demonstrate that our model outperforms alternative baselines, generating higher-quality complete designs. The generated human poses, framing and human image layout of our model are better aligned with the input design context, compared to those of the baselines.     

Our major contributions are threefold:
\begin{itemize}
    \item We formulate a new problem in the graphic design generation field: design-conditioned 3D human adding. The problem can be viewed as a specific form of human-aware design generation, where visual human representations need to be modeled explicitly. To the best of our knowledge, this is the first time that the two fields, 3D human generation and 2D graphic design generation, are bridged. We hope that our work can inspire future work on leveraging various 3D information for graphic design synthesis, which was previously addressed with 2D-only methods.
    \item We propose a model, POFRAY, which adds 3D humans into an input design by simultaneously generating 3D human poses, the 2D framing of the humans in an image, and the layout of the human image in the design.
    \item We build a benchmark, 3DHumanInDesign, to facilitate future research on design-conditioned 3D human adding.
\end{itemize}

\section{Relate Work}
\label{sec:Relatework}

\subsection{Graphic Design Generation}

Automatic graphic design generation has emerged as a crucial research area aimed at facilitating the design process.
CanvasVAE~\cite{yamaguchi2021canvasvae}  introduces a variational autoencoder to generate vector graphic designs represented as sets of canvas and element attributes. GOL~\cite{order} 
learns a design element ordering strategy to optimize the performance of design generative models. In recent years, a large body of methods on text-to-design generation has emerged to convert textual design intentions into multi-layer graphic designs, e.g., by building a cascaded system comprising multiple specialized models for different sub-tasks (e.g., background generation, typography generation)~\cite{cole,opencole},
formulating layered design generation as multi-layer transparent image generation~\cite{art}, 
utilizing a multimodal large language model to generate graphic designs as text-image documents~\cite{igd}, 
and training a simple diffusion model on sequences of discrete element attribute tokens ~\cite{LADEREIN}.There is also a line of work on automatic design composition, which aims to compose a set of input design elements into a complete design ~\cite{shabani2024visual,Graphist,LaDeCo}. 
FlexDM~\cite{flexdm} is a unified model that can solve multiple design tasks (e.g., image filling, layout generation, typography generation), by representing a graphic design as a set of elements, each comprising a set of fields, and formulating different design tasks as masked field prediction problems with different masking patterns.

Recent years have also seen a growing interest in developing models for generating layouts of graphic design elements~\cite{LayoutPrompter,jyothi2019layoutvae,Content-aware-Generative,VTN,PosterLlama,LayoutDiffusion,LayoutGAN,yang2026learninginteractionimagelayout}.
The success of Transformers in natural language processing has inspired their application to layout generation. LayoutTransformer~\cite{gupta2021layouttransformer} formulates layout generation as a sequence modeling problem and employs the self-attention mechanism to capture element relationships. BLT~\cite{kong2022blt} proposes a bidirectional Transformer that generates all layout tokens in parallel to enable flexible controllable generation.
Our method leverages Transformers to predict 3D human pose, 2D framing, and human image layout.

To our best knowledge, we are the first to model 3D human representations in 2D graphic designs, which has not been yet explored in previou studies.

\subsection{Human Pose Estimation and Generation}

3D Human pose and shape estimation
have been extensively studied in computer vision, with a range of applications in behavior analysis, animation and gaming.
Recent advances in expressive human pose and shape estimation (EHPS) enable the reconstruction of 3D whole-body meshes of a person\textemdash including the body, hands and face\textemdash from monocular images or videos using end-to-end models~\cite{OSX,cai2023smplerx,aios} 
that regress the parameters of a 3D parametric human model (e.g., SMPL-X~\cite{SMPLX}.

Conditional 3D human pose generation based on input conditions of various modalities has recently gained increasing attention. For example, ChatPose~\cite{feng2024chatpose} generates 3D human poses from natural language descriptions, optionally with image inputs. GenZI~\cite{li2024genzi} proposes a method to synthesize a posed 3D human that interacts with a given 3D scene according to an input text description. A recent study~\cite{jiang2024scaling} focuses on dynamic human-scene interaction modeling, generating 3D human motions (sequences of 3D poses) that are plausible in a given 3D scene context. NIFTY~\cite{kulkarni2024nifty} addresses the problem of object-conditioned human motion synthesis, which generates realistic 3D motions of a person interacting with a given 3D object.

In this work, we focus on generating 3D human poses in the context of graphic designs.

\section{Method}
\label{sec:method}

\begin{figure*}
  \centering
  \includegraphics[width=1\linewidth]{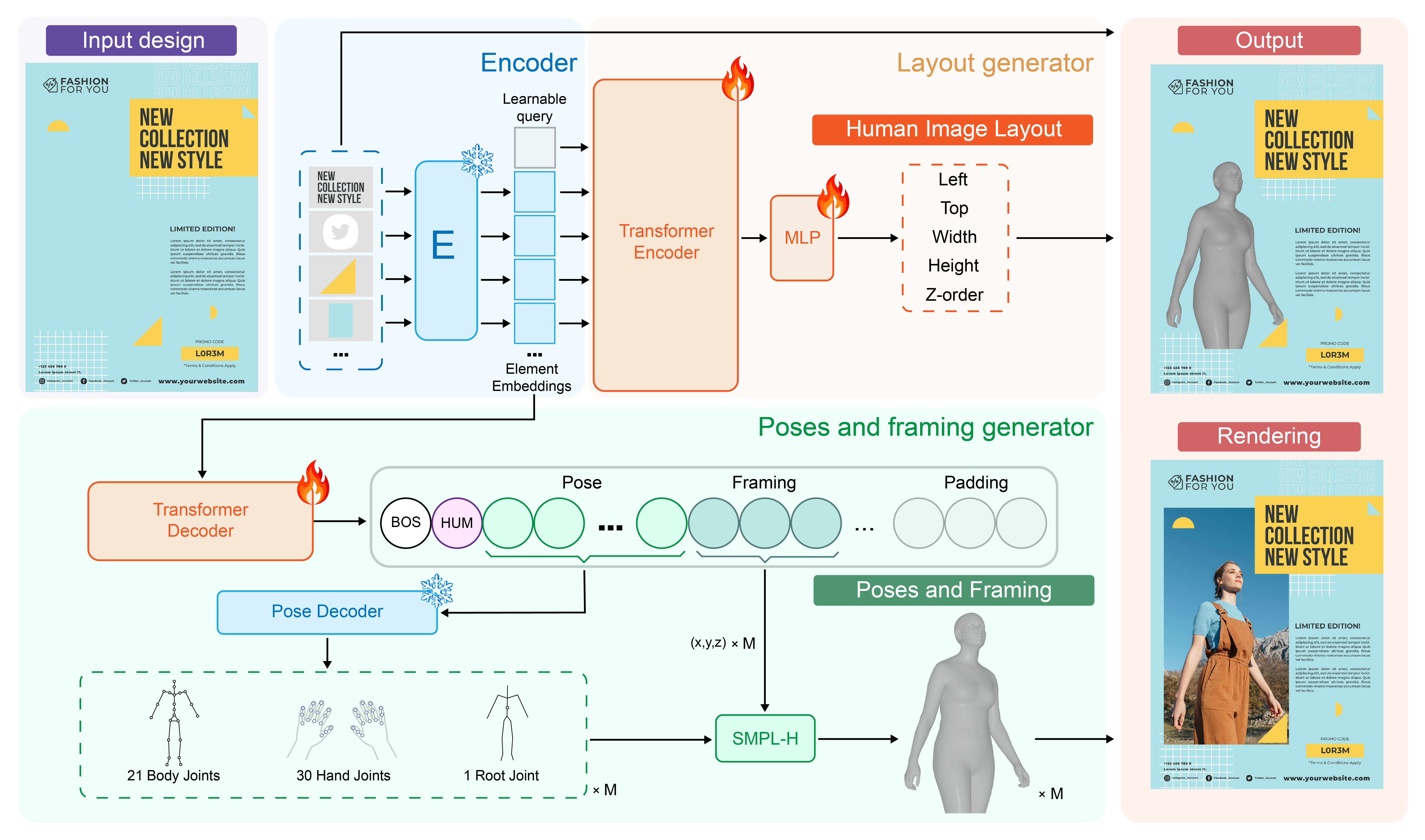}
  \caption{
  Overview of our model. It takes as input a partial graphic design with a set of multimodal elements (such as images, texts, vector shapes), and generates a complete design with one or several 3D humans added. An encoder is used first used to extract element-wise design embeddings, which are then fed into: a pose and framing generator, a layout generator. The pose and framing generator autoregressively generates pose tokens, which are then decoded into 3D human poses using a pre-trained pose decoder. It also predicts framing tokens that 
  control the composition of the 3D humans in an image. The layout generator takes as input a learnable query and all the element embeddings
  to predict the bounding box parameters and z-order of the human image in the design. With the predicted pose, framing and layout information, the 3D humans are composed into the design, and the 3D human representations can further be rendered as a realistic image.
  }
  \label{fig:model}
\end{figure*}

\subsection{Problem Formulation}
\label{sec:formulation}

Let $\mathcal{D} = \{e_1, e_2, \ldots, e_N\}$ denote an input graphic design consisting of $N$ design elements, where each element $e_i$ is described by a set of attributes: appearance attributes including 
font, color (font color or background color),
opacity; content 
(image or text content); layout attributes
$(c_i, l_i, t_i, w_i, h_i)$ specifying its type $c_i$, 
left position $l_i$, top position $t_i$, width $w_i$, and height $h_i$, respectively. $(l_i, t_i, w_i, h_i)$ are normalized bounding box parameters (with respect to the canvas) in the range $[0, 1]$.
Our objective is to build a model that takes as input a design $\mathcal{D}$, without a human image, and produces the following outputs:
\begin{itemize}
    \item \textbf{Pose}. A sequence of $M$ 3D human poses $P = \{\theta_1, \ldots, \theta_M\}$, where each pose is parameterized by the pose parameters $\theta_i \in \mathbb{R}^{52 \times 3}$ of the SMPL-H model~\cite{smplh},
    which specify 3D angular rotations for 52 joints (21 body joints, 30 hand joints, 1 root joint).
    \item \textbf{Framing}. Framing parameters $F = \{f_1, \ldots, f_M\}$ for $m$ 3D humans. The framing of human $j$ is parameterized by $f_j = (x_j, y_j, z_j)$, which is the global 3D position of the root joint (relative to the camera frame), which determines how the human is framed in the 2D image. 
    \item \textbf{Layout.} A human image layout $L = (l, t, w, h, z)$ that specifies the left position $l$, top position $t$, width $w$, height $h$, and z-order $z$ of the image containing the 3D humans in the design.
\end{itemize}

The generated poses $P$ are rendered into an image using the predicted framing parameters $F$, and placed into the input design according to the predicted image layout $L$, to create a cohesive human-included graphic design. 

\subsection{Our Model}
\label{sec:architecture}

As illustrated in Figure~\ref{fig:model}, our model consists of three main components: 1) a design encoder that extracts design context representations
from the input design, 2) a pose and framing generator that autoregressively generates human pose tokens and framing tokens 
, and 3) a layout generator that predicts the spatial arrangement of the human image. We now describe each component in detail.

\subsubsection{Design Encoder}
The design encoder follows the architecture of FlexDM ~\cite{flexdm}, projecting the elements in the input design into a set of element-wise embeddings.
In particular, for each element, 
the design encoder extracts three types of embeddings from its attributes:
\begin{itemize}
    \item \textit{Appearance embeddings.} We perform discretization on the font, color and opacity, and encode discrete font, color and opacity into separate embeddings 
    through learned embedding layers. 
    When an attribute is missing for an element (e.g., an image element does not have font color), an all-zeros embedding is used.

    \item \textit{Content embeddings.} The visual content is processed through the OpenCLIP~\cite{OpenCLIP} image encoder to obtain an image embedding, while the text content is encoded using OpenCLIP text encoder to produce a text embedding. Then, the image or text embedding is projected into a content embedding through a learned linear project layer.

    \item \textit{Layout embeddings.}
    Each of the four bounding box parameters is discretized. The element type and the discrete bounding box parameters are encoded into layout embeddings through learned embedding layers.
\end{itemize}

The embedding of the element is obtained by adding up its attribute embeddings.
All the element embeddings 
are then processed through a stack of transformer encoder blocks with self-attention layers
to capture inter-element relationships. The encoder finally outputs a sequence of contextualized design element embeddings, 
which we refer to as 
\textit{a design context representation} $\mathbf{C}$ that will be used to condition subsequent generators on information in the input partial design. 

\subsubsection{Pose and Framing Generator}

To generate human poses, we adopt a discrete pose representation inspired by TokenHMR~\cite{tokenhmr}. We first train a Vector Quantized Variational Autoencoder (VQ-VAE) tokenizer specifically for the SMPL-H pose parameters.
The VQ-VAE encoder maps continuous SMPL-H pose parameters $\theta \in \mathbb{R}^{52 \times 3}$ to a discrete token sequence $\mathbf{z}_p = (z_1, z_2, \ldots, z_{400})$, where each token $z_i$ corresponds to a code vector in a learned codebook of size $64$. The VQ-VAE decoder can reconstruct the pose parameters from the tokens: $\hat{\theta} = \mathcal{D}_\text{pose}(\mathbf{z}_p)$.
For framing parameters $f = (x, y, z)$, we discretize each coordinate into 256 bins by applying the K-means clustering on the training set, resulting in a sequence of framing tokens $\mathbf{z}_f = (z_x, z_y, z_z)$.

For $m$ humans, their poses and framing configurations can then be represented as a sequence of discrete tokens:
\begin{equation}
    \mathbf{Z} = (\text{BOS}, \text{HUM}, \mathbf{z}_{p_1}, \mathbf{z}_{f_1}, \ldots, \mathbf{z}_{p_m}, \mathbf{z}_{f_m}, \text{EOS})
\end{equation}
where BOS and EOS are special tokens denoting the start and end of the sequence, and the HUM token is used to signals the start of a human.

Our pose decoder employs an
autoregressive Transformer architecture trained to generate tokens in the sequence $\mathbf{Z}$ iteratively conditioned on the design context representation $\mathbf{C}$ from the design encoder. 
The design context representation is injected into the pose decoder through cross-attention layers. 
The generated pose token sequence $\hat{\mathbf{z}}_p$ is decoded back to the SMPL-H pose parameters using the pre-trained VQ-VAE decoder $\mathcal{D}_\text{pose}$, and the generated framing tokens are converted to continuous framing parameters by taking the bin center values of the tokens.

\subsubsection{Layout Generator}

The layout decoder predicts the spatial arrangement of the human image within the design. Our layout decoder utilizes a bidirectional Transformer architecture (with a stack of Transformer encoder blocks), whose inputs include a learnable query (a learnable embedding) and the design context representation $\mathbf{C}$ (a sequence of element embeddings). The query attends to all the element embeddings through the self-attention layers to extract an output query representation (the output embedding of the last Transformer block for the query) that contains necessary information for layout prediction. The output query representation is finally decoded into the layout parameters $(\hat{l}, \hat{t}, \hat{w}, \hat{h}, \hat{z})$ with a  two-layer MLP.

\subsubsection{Human Image Generation and Composition}
Given the generated pose, framing and layout parameters, we can create a human image and compose it into the input design. Specifically, we feed the pose and framing parameters into the SMPL-H model to generate posed 3D human meshes placed in the camera coordinate system, and project the meshes onto an image of size $(W \cdot \hat{w}, H \cdot \hat{h})$, where $(W, H)$ is the canvas size of the input design and $(\hat{w}, \hat{h})$ is the predicted human image size. This results in a transparent human mesh image (where the alpha values of pixels outside the human regions are set to 0), which can be composed into the input design according to the predicted position and z-order $(\hat{l}, \hat{t}, \hat{z})$. To generate a realistic human image, we create a 2D pose map using the 2D projections of the 3D joint positions of the human meshes, and employ LayerDiffuse ~\cite{layerdiffuse} (with SDXL ~\cite{podell2023sdxl} to generate a transparent image conditioned on the pose map, the input design, and a text prompt. The text prompt is generated by Qwen3-VL~\cite{Qwen3-VL} that is queried to describe the human appearance according to the input design. The pose map is incorporated into the SDXL through ControlNet~\cite{controlnet}; the input design is sent into the SDXL through IP-Adapter~\cite{ye2023ipadaptertextcompatibleimage}.

\subsection{Training}
\label{sec:training}

Our model is trained with two task-specific loss functions:

\begin{itemize}
    \item \textit{Pose and framing loss.} The pose and framing generator is trained with a cross-entropy loss for next-token prediction. For a sequence $\mathbf{Z}$ of length $L$. The loss is defined as:
    \begin{equation}
        \mathcal{L}_{\text{PF}} = -\sum_{t=2}^{L} \log p_{\phi}(\mathbf{Z}_t | \mathbf{Z}_{<t}, \mathbf{C}).
    \end{equation}

    \item \textit{Layout loss.} The layout generator is supervised with a combination of a L1 loss and a Generalized IoU (GIoU) loss:
    \begin{equation}
        \mathcal{L}_{\text{L}} = \mathcal{L}_{L1} + \mathcal{L}_{\text{GIoU}}.
    \end{equation}
    $\mathcal{L}_{L_1} = \left| \hat{L} - L \right|$, where $\hat{L}$ and $L$ are the predicted and ground truth human image layout parameters, respectively. $ \mathcal{L}_{\text{GIoU}} = 1 - \text{GIoU}(\hat{B}_{h}, B_{h})$, where $\hat{B}$ and $B$ are the predicted and ground truth bounding boxes of the human image, respectively.

\end{itemize}

During training, we first fine-tune FlexDM on the training component of our dataset (Section \ref{sec:dataset}) and use the obtained corresponding weights to initialize our design encoder. 
Then, we freeze the design encoder weights, and train our model via a three-stage training procedure:
\begin{itemize}
    \item \textit{Stage 1: Pose and framing generator pre-training.} We first train 
    the pose and framing generator using the loss $\mathcal{L}_{\text{PF}}$. This stage allows the pose and framing generator to capture the dependency of human pose and framing on design context. 

    \item \textit{Stage 2: Layout generator pre-training.} We train the layout generator using the loss $\mathcal{L}_{\text{L}}$. This stage enables the layout generator to learn spatial placement of human images based on design context features.

    \item \textit{Stage 3: Joint Fine-tuning.} Finally, we jointly fine-tune the two generators, with a total loss:
    \begin{equation}
        \mathcal{L}_{\text{Total}} = \mathcal{L}_{\text{PF}} + \mathcal{L}_{\text{L}}.
    \end{equation}
    This joint fine-tuning stage allows the two generators to produce the pose, framing and layout that are harmonious with each other.
\end{itemize}

\section{Experiments}

\subsection{Dataset}
\label{sec:dataset}

Our \textit{3DHumanInDesign} dataset consists of $\sim$10k \textit{layered} graphic designs with human images, carefully curated from two sources to ensure diversity and quality. Specifically, we collect 4,479 samples from the \textit{Crello} dataset~\cite{yamaguchi2021canvasvae}, which covers different kinds of graphic designs such as web banners, social media posts and posters.
Additionally, we gather 5,460 magazine covers and posters from \textit{Freepik}\footnote{https://www.freepik.com}.
For the human image on each design, we extract SMPL-H pose parameters and 3D root joint positions (framing parameters) 
using a previous human pose and shape estimation method~\cite{aios}. 
We remove human images from the original designs to create input designs (without human images).
We partition our dataset into training and test splits with a ratio of 8:2, while ensuring that each split has $40 \%$ and $60 \%$ of its samples from Crello and Freepik, respectively.

\subsection{Implementation Details}

For all the three training stages, we optimize our model using AdamW ($\beta_1 = 0.9$, $\beta_2 = 0.95$, weight decay $0.1$) with a base learning rate of $1\times10^{-4}$, using $1{,}000$ warm-up iterations followed by a cosine decay schedule. We use a batch size of $16$, a gradient-norm clip of $1.0$, a label smoothing of $0.05$, and maintain an exponential moving average of the weights with a decay of $0.995$.At inference time, We employ top-$p$ sampling with temperature $0.8$ and $p = 0.8$.

\subsection{Baselines}

Since there is no prior work on design-conditioned 3D human adding, we compare our method against two baselines:

\begin{itemize}
    \item \textit{FlexDM}. FlexDM~\cite{flexdm} is a unified model for solving different design tasks.
    We fine-tune FlexDM on our training set, 
    and adapt it to our problem by empolying a two-stage approach. In the first stage, we utilize FlexDM to solve an image filling task to produce an image embedding, by feeding all the elements in the input design, along with an added image element (with all the fields masked out except its category), into FlexDM. The output image embedding is used to retrieve an existing image from a database of realistic human images (taken from the designs in the test set) 
    via cosine similarity between image embeddings. Then, the pose and framing parameters are estimated from the retrieve image using an off-the-shelf human pose and shape estimation method~\cite{aios}.
    In the second stage, we use FlexDM to perform layout generation for the retrieved image.
    Since FlexDM can not predict the z-order of an element, for a fair comparison, we use the z-order generated by our model when composing the human image into the input design. 

    \item \textit{MLLM.} We design a baseline leveraging multimodal large language models (MLLM) for 
    design-conditioned 3D human adding.
    Specifically, we provide the input design image along with its layout information (in textual format) to Gemini 3 Pro~\cite{geminiteam2023gemini}, and prompt it to predict: 1) a text description for a human image; 2) the bounding box parameters and z-order of the human image in the design. The image description is then fed into a text-to-image model, Nano Banana Pro~\cite{google2025nanobanana}, to generate a realistic human image.
    Finally, the pose and framing parameters are extracted from the generated image using an off-the-shelf human pose and shape estimation method~\cite{aios}.
\end{itemize}

\subsection{Evaluation Metrics}
We quantitatively evaluate the performance of our method using four metrics that cover complementary aspects: human-design coherence, overall design quality, pose accuracy, and layout quality.

\begin{itemize}
    \item \textit{Human-Design Alignment (HD-Align).} Inspired by CLIP~\cite{clip}, we train a contrastive model to learn a joint embedding space of human representations and partial designs (with human images excluded), where matched human and design lie close and unmatched ones fall far apart. 
    Our human representation combines the 3D pose, framing, and human image layout information.
    We compute cosine similarity between human and design embeddings as a HD-Align score, with higher HD-Align indicating better human-design coherence.
    \item \textit{Design FID (D-FID).} We train a variational autoencoder that encodes both a human representation (combining 3D pose, framing, and human image layout information) and a partial design (without any human image) into a holistic design embedding. We use these embeddings to compute Fr\'{e}chet Inception Distance (FID) between generated and real designs as a measure of overall design quality. Lower D-FID indicates better holistic design quality.
    \item \textit{MPJPE.} We adopt Mean Per Joint Position Error (MPJPE), a common metric for 3D human pose estimation, 
    to evaluate predicted 3D poses,
    which computes the mean Euclidean distance between predicted and ground-truth 3D joint positions.
    \item \textit{Alignment, Overlap.} We evaluate the quality of predicted human image layouts (bounding boxes) with two widely used metrics in the layout generation literature, \textit{Alignment} and \textit{Overlap}~\cite{LayoutGAN}. Both metrics are computed on the layouts of complete designs with human images included. 
\end{itemize}

\subsection{Quantitative Results}

\begin{table}[t]
\centering
\caption{Quantitative evaluation results of different methods. The best results are in bold, while the second best results are underlined.}
\label{tab:quantitative_results}
\resizebox{\columnwidth}{!}{%
\begin{tabular}{l|ccccc}
\hline
Method
& HD-Align~$\uparrow$ & D-FID~$\downarrow$  
& MPJPE~$\downarrow$ 
& Align~$\downarrow$ & Overlap~$\downarrow$ \\
\hline
FlexDM     & 4.65 & \underline{3.90}& \underline{2176.29} & \underline{0.17} & 213.39 \\
MLLM       & \underline{5.26} & 7.95 & 2829.97 & 0.20 & \textbf{196.03} \\
Ours       & \textbf{6.52} & \textbf{0.07} & \textbf{385.12} & \textbf{0.17} & \underline{209.11} \\
\hline
Real Data  & 7.06 & -- & -- & 0.16 & 210.48 \\
\hline
\end{tabular}%
}
\end{table}

Table~\ref{tab:quantitative_results} reports the quantitative results of different methods on the test set. The HD-Align score of our method is better than those of the baselines by significant margins, while being competitive with that of the read data (samples in the test set). 
This suggest great alignment of our generated human pose, framing and human image layout with the design context. Our method also shows a noticeable improvement in D-FID compared to the other methods, highlighting the superior quality of holistic designs (with added human images) generated by our method. Our method significantly outperforms the baselines in MPJPE, confirming that its generated poses are closer to the ground truth. In addition, our method achieves the best alignment score and the second best overlap score, which means that our method can result in spatially cohesive design layouts after adding human images into the partial designs. Although MLLM yields the lowest overlap score, the overlap score of our method (209.11) is quite close to that of the real data (210.48), which suggests that our resulting layouts encompass a similar amount of overlap between elements to real designs.   

\begin{figure*}[t]
  \centering
  \includegraphics[width=0.9\linewidth]{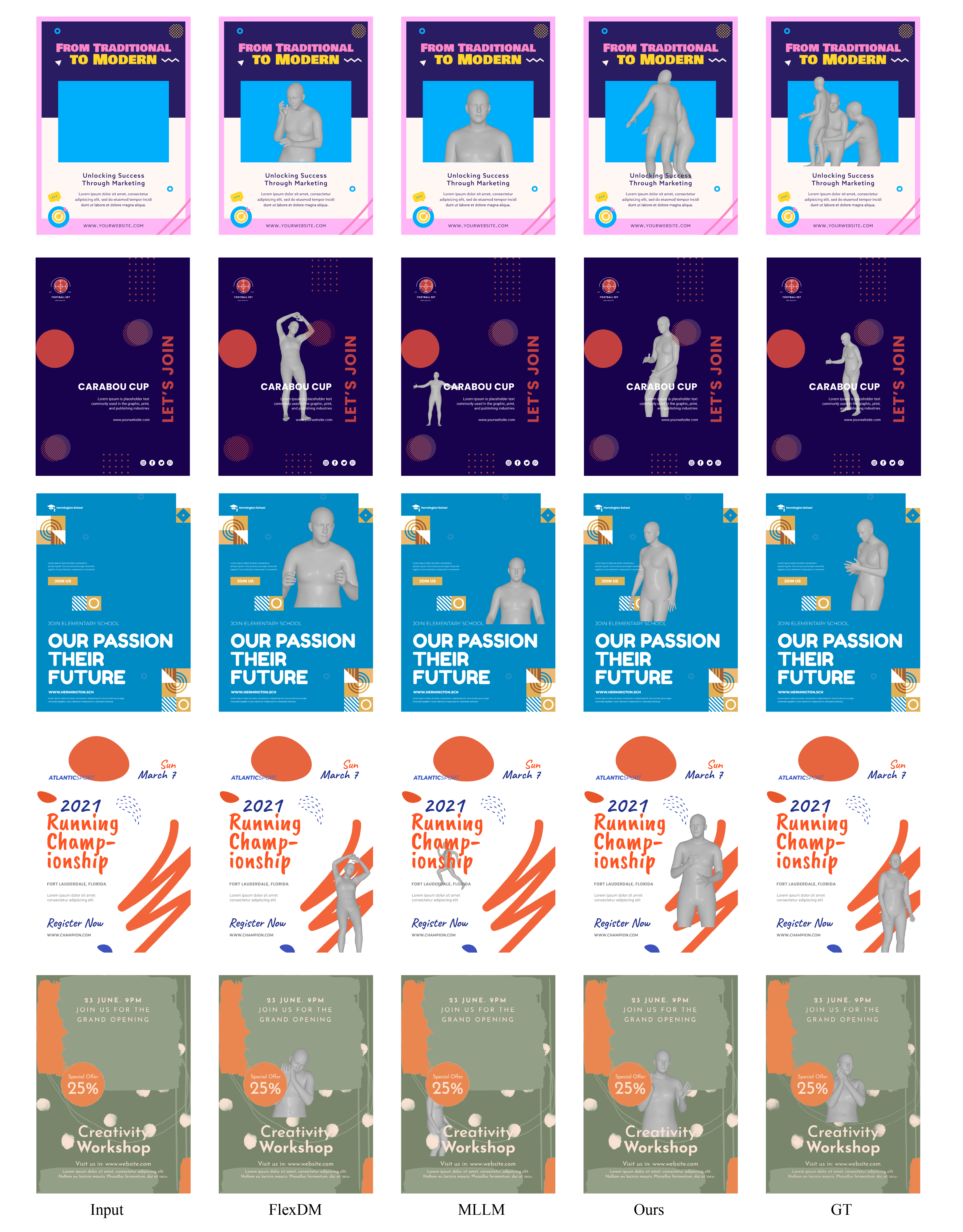}
  \caption{
  Qualitative comparison of different methods.
  }
  \label{fig:comparison}
\end{figure*}

\begin{figure*}[t]
  \centering
  \includegraphics[width=0.85\linewidth]{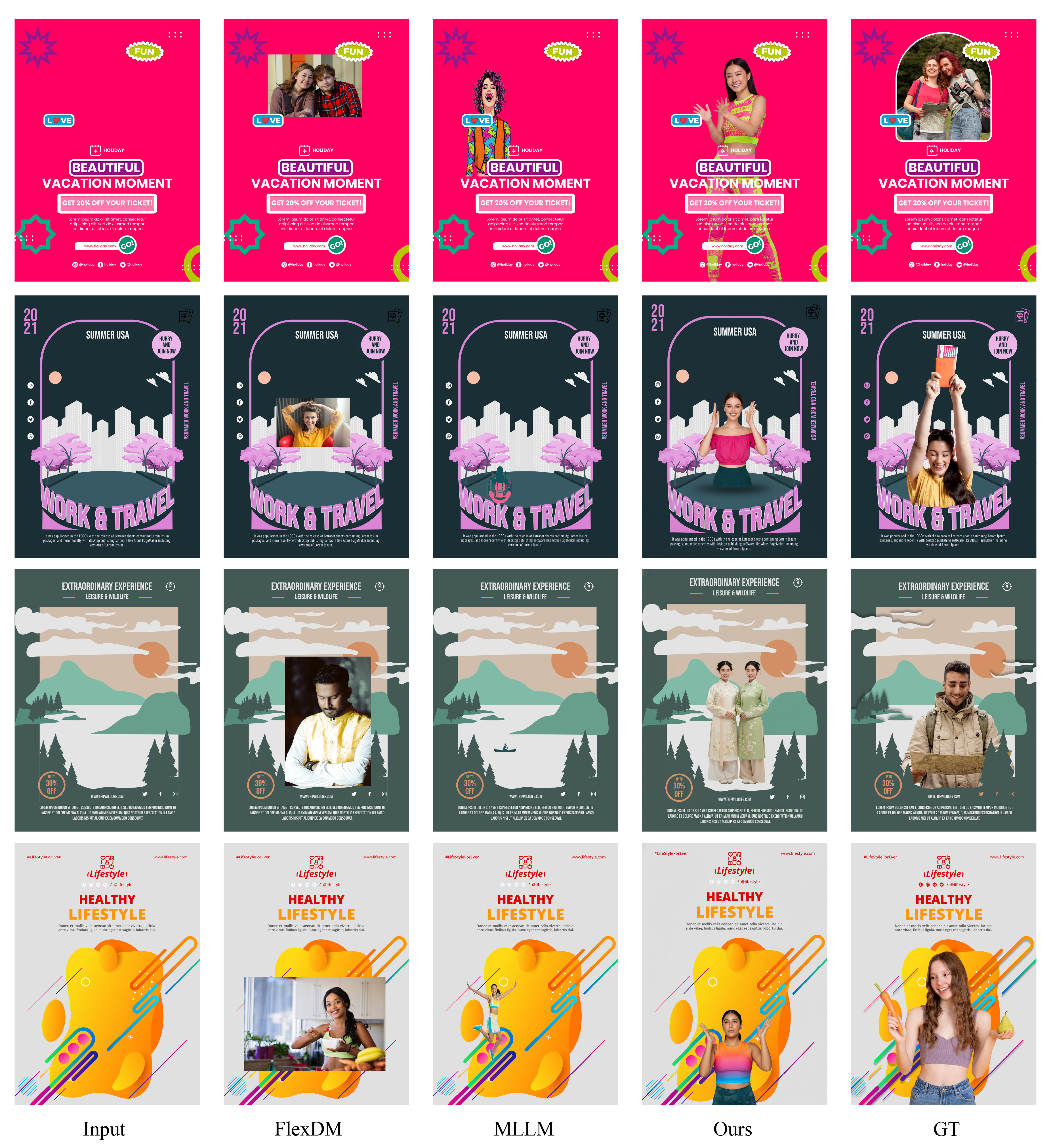}
  \caption{Results of different methods where realistic human images are added.
  }
  \label{fig:rendering}
\end{figure*}

\subsection{Qualitative Results}

Figure~\ref{fig:comparison} shows visual comparison of results from different methods. As compared with the baselines, our method can generate 3D human poses and 2D framing that better harmonizes with the design context and more closely resemble what professional designers create. For example, in the first row, our method add multiple humans whose \textit{body orientations} are similar to those of humans in the ground truth design, whereas FlexDM and MLLM generates a single human facing outwards. In the third row, our method frames the added human in the same way as the real design\textemdash a \textit{medium full shot} is used, capturing the human from the top of the head to just below the waist, while FlexDM and MLLM employ a \textit{medium shot} that starts just above the head and ends around the waist. In addition, our method is able to generate human image layouts that are spatially compatible with the remaining design elements. MLLM sometimes falls shot of predicting proper bounding box coordinates and z-order for the added human image, causing visual conflict between the human image and other design elements; for example, a large portion of the added human is occluded by the texts in the fourth row.

\subsection{User Study}

\begin{table}[t]
\centering
\caption{User study results. For each aspect, the percentage of the time that each method is chosen as the best is reported.
}
\label{tab:user_study}
\resizebox{0.85\columnwidth}{!}
{
\begin{tabular}{l|cc}
\hline
 Method & Human-Design Align & Aesthetic Quality\\
\hline
FlexDM  & 11.9\% & 17.8\% \\
MLLM &  27.0\% & 22.5\% \\
Ours & \textbf{61.1\%} & \textbf{59.7\%} \\
\hline
\end{tabular}
}
\end{table}

We conduct a user study comparing our method with FlexDM and MLLM for \textit{human-design alignment} and \textit{aesthetic quality}. We apply each method to 100 input designs from our test set to generate complete designs (with realistic human images added). Our study involves 23 professional graphic designers. The participants are shown three generated complete designs by different methods in randomized order, along with the input design and ground truth design for reference, and asked to evaluate the generated designs in terms of two aspects: 1) human-design alignment (whether the added human image fits well the remaining part of the design in terms of human pose, framing and image layout; 2) aesthetic quality (whether the design looks visually aesthetic). For each aspect, the participants are asked to choose the best generated design. The results are shown in Table ~\ref{tab:user_study}. Our method is significantly preferred over the other methods on both aspects, suggesting that our method is able to achieve better human-design alignment and aesthetic quality. Figure \ref{fig:rendering} shows some results of different methods that are displayed in the user study.
\section{Conclusion}

In this work, we make the first step to bring 3D human modeling into the creation of 2D graphic designs, by formulating and investigating the problem of design-conditioned 3D human adding, where we aim to add 3D human representations into a partial input design. Our experimental results demonstrate the superiority of our proposed model in generating the 3D human poses, 2D framing and human image layout that better fit with the remaining elements in the input design, compared to several baselines. We believe that our effort in this work can inspire future work in building 3D-aware generative models of graphic design. 

\newpage

\bibliographystyle{ACM-Reference-Format}
\bibliography{references}

\appendix

\end{document}